\documentclass[a4paper,11pt]{article}

\pdfoutput=0 

\usepackage{jheppub} 

\usepackage[T1]{fontenc} 

\def\as{\alpha_s}
\def\d{{\rm d}}

\def\lr{\left( }
\def\rr{\right) }
\def\le{\left[ }
\def\re{\right] }

\def\beq{\begin{equation}}
\def\eeq{\end{equation}}
\def\bea{\begin{eqnarray}}
\def\eea{\end{eqnarray}}

\title{\boldmath NLO Monte Carlo predictions for heavy-quark production at the LHC:
pp collisions in ALICE}

\author[a]{M.\ Klasen,}
\author[b,c]{C.\ Klein-B\"osing,}
\author[a]{K.\ Kovarik,}
\author[d]{G.\ Kramer,}
\author[a]{M.\ Topp}
\author[b]{and J.\ P.\ Wessels}

\affiliation[a]{Institut f\"ur Theoretische Physik, Westf\"alische
 Wilhelms-Universit\"at M\"unster, \\ Wilhelm-Klemm-Stra\ss{}e 9,
 D-48149 M\"unster, Germany}
\affiliation[b]{Institut f\"ur Kernphysik, Westf\"alische
 Wilhelms-Universit\"at M\"unster, \\ Wilhelm-Klemm-Stra\ss{}e 9,
 D-48149 M\"unster, Germany}
\affiliation[c]{
ExtreMe Matter Institute, GSI,
Planckstra\ss{}e 1, D-64291 Darmstadt,
Germany}
\affiliation[d]{
II.\ Institut f\"ur Theoretische Physik, Universit\"at Hamburg, Luruper
Chaussee 149, D-22761 Hamburg, Germany}

\emailAdd{michael.klasen@uni-muenster.de}
\emailAdd{c.klein-boesing@uni-muenster.de}
\emailAdd{karol.kovarik@uni-muenster.de}
\emailAdd{kramer@mail.desy.de}
\emailAdd{m$\_$topp02@uni-muenster.de}
\emailAdd{j.wessels@uni-muenster.de}

\preprint{DESY 14-074, MS-TP-14-21}

\abstract{
Next-to-leading order (NLO) QCD predictions for the production of heavy quarks
in proton-proton collisions are presented within three different approaches to
quark mass, resummation and fragmentation effects. In particular, new NLO and parton
shower simulations with POWHEG are performed in the ALICE kinematic regime
at three different centre-of-mass energies, including scale and parton density variations,
in order to establish a reliable baseline for future detailed studies of heavy-quark
suppression in heavy-ion collisions.
Very good agreement of POWHEG is found with FONLL, in particular for centrally produced
$D^0$, $D^+$ and $D^{*+}$ mesons and electrons from charm and bottom quark decays,
but also with the generally somewhat higher GM-VFNS predictions within the theoretical
uncertainties. The latter are dominated by scale rather than quark mass variations. Parton
density uncertainties for charm and bottom quark production are computed here with POWHEG
for the first time and shown to be dominant in the forward regime, e.g.\ for muons
coming from heavy-flavour decays. The fragmentation into $D_s^+$ mesons
seems to require further tuning within the NLO Monte Carlo approach.}

\begin{document} 
\maketitle
\flushbottom

\section{Introduction}
\label{sec:1}

The production of heavy quarks in hadronic collisions is an important probe
for various aspects of QCD. In proton-proton collisions, this process is
perturbatively calculable down to low transverse momenta ($p_T$) due to the
presence of the heavy-quark mass $m$. Calculations can there be performed
reliably in the Fixed Flavour Number Scheme (FFNS), where the heavy flavour
appears only in the hard scattering process. Conversely, when $p_T\gg m$ the
Zero-Mass Variable Flavour Number Scheme (ZM-VFNS) applies, where the heavy
quark appears also as an active parton in the initial-state Parton Density
Functions (PDFs) and final-state Fragmentation Functions (FFs) and its mass
can be neglected in the hard scattering process. The intermediate regime
requires delicate matching procedures, that have been developed under the
names of Fixed Order plus Next-to-Leading Logarithms (FONLL)
\cite{Cacciari:1998it} and General-Mass Variable Flavour Number Scheme
(GM-VFNS) \cite{Kniehl:2004fy}.

While these calculations are mostly analytic and provide only an accurate
description of the inclusively produced heavy hadron, the matching of
next-to-leading order (NLO) calculations with parton showers in MC@NLO
\cite{Frixione:2003ei} or POWHEG \cite{Frixione:2007nw} allows to provide
a more complete description of the hadronic final state. It is modeled by
parton showers and string or cluster fragmentation and can be subject to
decays and even detector response, while at the same time NLO QCD accuracy
in the hard and improved leading-logarithm resummation in the soft/collinear
regimes are retained. In this respect these predictions are therefore
superior to those from general-purpose Monte Carlo generators like PYTHIA
\cite{Sjostrand:2007gs} or HERWIG \cite{Corcella:2000bw}, which also provide
a complete hadronic final state, but only with leading order accuracy.

In this paper we compare the three different theoretical approaches mentioned
above in the kinematic regime relevant for the ALICE experiment at the LHC.
The goal is to establish the reliability of the proton-proton baseline for
future studies of heavy-ion collisions, where the suppression (quenching)
of heavy quarks at large transverse momenta is an important signal for the
deconfined state of matter, the so-called quark-gluon plasma (QGP). To this
end, we compute, with FONLL, GM-VFNS and POWHEG, the production of $D^0$, $D^+$,
$D^{*+}$, and $D_s^+$ mesons in pp collisions at $\sqrt{s}=2.76$ and 7 TeV as well
as the production of muons and electrons originating from heavy-flavour (charm
and bottom quark) decays at these energies. For the latter, we also make
predictions at a centre-of-mass energy
of $\sqrt{s}=5.02$ TeV, pertinent to the 2013 pPb run of the LHC whose data are currently
being analysed by the experimental collaborations. We also assess
the theoretical uncertainties coming from
independent variations of the renormalisation and factorisation scales, the heavy-quark
mass, and
from the complete error set of CTEQ6.6 parton densities \cite{Nadolsky:2008zw}.
Factorisation scheme uncertainties can in principle also be important
\cite{Klasen:1996yk}, but are usually not considerered.

Previously,
only the FONLL and GM-VFNS calculations had been compared with each other for
$D^0$, $D^+$ and $D^{*+}$ meson production at $\sqrt{s}=2.76$ TeV
\cite{Abelev:2012vra} and 7 TeV \cite{ALICE:2011aa} and for $D_s^+$ meson production
also at 7 TeV \cite{Abelev:2012tca}. On the other hand, MC@NLO had  been
tested against NLO in the FFNS \cite{Frixione:2003ei} and POWHEG against MC@NLO
\cite{Frixione:2007nw} for top and bottom quark production at the Tevatron
and at the nominal LHC energy of $\sqrt{s}=14$ TeV. More recently, FONLL, MC@NLO
and POWHEG predictions have been compared for $D^+$ and $B^+$ meson production at an
energy of 7 TeV  \cite{Cacciari:2012ny}, but no predictions were made for other
charm mesons or at the centre-of-mass energies of 2.76 and 5.02 TeV. Furthermore, a
systematic scale and PDF uncertainty study has only been performed there for FONLL
and not for POWHEG, and no comparisons to the GM-VFNS have been made.
Attempts to describe the quenching of heavy quarks are mostly and at least in principle
based on QCD, but
they range in practice from the consideration of radiative energy loss supplemented
with in-medium dissociation \cite{He:2011pd} 
or collisional energy loss \cite{Horowitz:2011cv,Buzzatti:2011vt,Gossiaux:2009mk,Gossiaux:2010yx} 
over a Boltzmann approach to multi-parton scatterings \cite{Fochler:2011en} 
to the relativistic Langevin equation with multiple uncorrelated random collisions \cite{Alberico:2011zy,Monteno:2011gq} 
and AdS/CFT drag coefficients \cite{Horowitz:2011wm}. 
Note, however, that
the description of heavy-quark quenching is beyond the scope of our present work.

This paper is organised as follows: In Sec.\ \ref{sec:2} we describe the
relevant features of the three different theoretical approaches (FONLL, GM-VFNS
and POWHEG) that we employ and compare in this work. Sec.\ \ref{sec:3} contains our main
numerical results, i.e.\ central predictions and associated theoretical uncertainties for
transverse-momentum distributions of heavy quarks produced in the ALICE kinematic regime
at the LHC with three different centre-of-mass energies. We  summarise our results
and present our conclusions in Sec.\ \ref{sec:4}.

\section{Theoretical input}
\label{sec:2}

In this section, we describe the main features of the three different theoretical
approaches to heavy-quark production (FONLL, GM-VFNS and POWHEG), that will later
form the bases of our numerical predictions for pp collisions at various centre-of-mass energies
of the LHC. The most important parameter is of course the heavy-quark mass $m$, which
we take to be 1.5 GeV for charm and 4.75 GeV for bottom quarks. These are the default
values in both FONLL and POWHEG and have in particular been used to obtain the central
FONLL predictions in the ALICE publications \cite{Abelev:2012vra,ALICE:2011aa,%
Abelev:2012tca,Abelev:2012qh,Abelev:2012xe,Abelev:2012sca,Abelev:2012pi}.

For the same reasons and in line with common practice, we employ as our default central
renormalisation and factorisation scales the square root of the
quadratic mean of the heavy-quark mass $m$
and the transverse momentum $p_T$ multiplied by $\sqrt{2}$ and set $\mu=\mu_f=\sqrt{m^2+p_T^2}$.
This scale is also used as
the default value of the final-state fragmentation scale $\mu_D$, which in GM-VFNS predictions
can be varied independently of the initial-state factorisation scale $\mu_f$. Independent
variations of these scales will be performed by factors of two about the central value,
but omitting the extreme combinations that amount to factors of four between the different
scales.

As our central PDF set we employ the CTEQ6.6 parametrisation \cite{Nadolsky:2008zw}, again
because it was already used in the FONLL calculations shown in the ALICE publications.
Note, however, that these PDFs have been obtained with $m_c=1.3$ GeV and $m_b=4.5$ GeV, which
are slightly lower than our default quark mass values. This influences in particular the $c$ and $b$
quark PDFs, which enter the evolution equations with zero input at the starting scales $\mu_0
=m_c$ and $m_b$. The value of the QCD scale in CTEQ6.6 in the $\overline{\rm MS}$ scheme and
for five light quark flavors is $\Lambda^{n_f=5}_{\overline{\rm MS}}=226$ MeV, which gives
$\alpha_s(M_Z)=0.1181$ at $M_Z=91.1876$ GeV in agreement with the world average of
$\alpha_s(M_Z)=0.1184\pm0.0007$ \cite{Beringer:1900zz}.
The PDF uncertainty will be estimated in POWHEG using the usual formulas
\bea
 \delta^+f&=&\sqrt{\sum_{i=1}^{22}[\max(f_i^{(+)}-f_0,f_i^{(-)}-f_0,0)]^2},
 \label{eq:a}\\
 \delta^-f&=&\sqrt{\sum_{i=1}^{22}[\max(f_0-f_i^{(+)},f_0-f_i^{(-)},0)]^2},
 \label{eq:b}
\eea
where $f_i^\pm$ are the PDFs for positive and negative variations of the
PDF parameters along the $i$-th eigenvector direction in the 22-dimensional
PDF parameter space.

In the following, we turn to the description of the specific theoretical assumputions
entering the FONLL, GM-VFNS and POWHEG approaches.

\subsection{FONLL}

The Fixed-Order plus Next-to-Leading-Logarithms (FONLL) calculation of Cacciari et al.\
\cite{Cacciari:1998it} is based on the matching of NLO massive and massless calculations
according to the prescription
\bea
 \d\sigma_{\rm FONLL}&=&\d\sigma_{\rm FO}+(\d\sigma_{\rm RS}-\d\sigma_{\rm FOM0})\times G(m,p_T).
\eea
Here, FO denotes the massive NLO cross section, where the heavy-quark flavour $h$
enters only in the partonic scattering through so-called flavour creation processes,
but not in the PDFs, and its mass is kept as a non-vanishing parameter. The
NLO partonic cross section then includes mass logarithms of the form
\bea
 \d\sigma_{\rm FO}&\propto&{\alpha_sC_F\over2\pi}\lr{1+z^2\over1-z}\rr_+ \log{p_T^2\over m^2}+\dots,
\eea
where $m$ regularises the collinear singularity, e.g.\ of the splitting $h\to hg$,
and where the logarithm becomes very large when $p_T\gg m$, thus spoiling the
perturbative expansion in $\alpha_s$. This part, which is singular in the massless
limit ($m\to 0$), and the finite parts related to its different definition in
dimensional and mass regularisation are denoted FOM0 and therefore resummed to NLL
order in the contribution denoted RS. RS is then added to the FO calculation, while
the overlap FOM0 is subtracted to avoid double counting.

The resummation relies on the perturbative FFs for the probability of a heavy quark $h$,
gluon, or light parton $i$ to go into a heavy quark $h$ \cite{Mele:1990cw}, which satisfy
Altarelli-Parisi evolution equations. Their initial values at the starting scale $\mu_0$
are calculable perturbatively. In the $\overline{\rm MS}$ scheme they are given by
\bea
 D_h(z,\mu_0)&=&\delta(1-z)+{\alpha_s(\mu_0)C_F\over2\pi}\le{1+z^2\over1-z}\lr \log{\mu_0^2\over m^2}+2\log(1-z)-1 \rr\re_+,\\
 D_g(z,\mu_0)&=&{\alpha_s(\mu_0)T_F\over2\pi}(z^2+(1-z)^2)\log{\mu_0^2\over m^2}, ~~{\rm and}\\
 D_i(z,\mu_0)&=&0, \qquad {\rm for}\ i\neq g,h,
\eea
respectively. The second term relates to the gluon-splitting contribution $g\to h\bar{h}$, while
the third term describes the coupling of light to heavy quarks. It occurs only at
next-to-next-to-leading order through $i\to ig\to ih\bar{h}$ and is therefore neglected.
As usual, $C_F=4/3$ and $T_F=1/2$.

The perturbative FFs are evolved to the factorisation scale $\mu_f$ and convoluted
with the NLO cross sections for massless partons, subtracted in the $\overline{\rm MS}$
scheme, so that also so-called flavour excitation
processes are included. These involve the heavy quark also as an active parton in the PDFs.
The result is then convoluted with non-perturbative functions to describe the hadronisation
of heavy quarks into heavy hadrons.
For bottom quarks, the functional form
\bea
 D_b(z)&=&(\alpha+1)(\alpha+2)z^\alpha(1-z)
\eea
is used, which is normalised to unity and where the single parameter $\alpha=24.2$
has been fitted to LEP data relevant to the production of a mixture of $B$ hadrons,
since no data are available for individual hadrons like $B^0$ or $B^+$ \cite{Cacciari:2005uk}.
For charm quarks, experimental data for individual mesons ($D^0$, $D^+$, $D^{*}$, and
$D_s^+$) exist. They have been exploited, together with the theoretical understanding
of the similarities and differences in the fragmentation of a heavy quark into
pseudoscalar ($D$) and vector ($D^*$) mesons, to
construct non-perturbative functions for $D$ mesons exclusively in terms of $c\to D^*$
fragmentation. Its single parameter $r=0.1$, relating $D$ to $D^*$ mesons,
was extracted from LEP data, together with the
different branching ratios, and also used to describe primary pseudoscalar $D$ meson production
\cite{Cacciari:2003zu}.
The fragmentation is numerically performed by rescaling the quark three-momentum
at a constant angle in the laboratory frame.
The decay of the $D$ and $B$ mesons into electrons (or muons) is controlled by the
experimentally measured decay spectra using data from the BaBar and CLEO
collaborations \cite{Cacciari:2005rk}, normalised to branching ratios at high energy
\cite{Beringer:1900zz}.

The matching function $G(m,p_T)$ must tend to unity in the massless limit
$p_T\gg m$, where FO approaches FOM0 and the mass logarithms must be resummed.
However, in FONLL its functional form is not simply unity, but rather
\bea
 G(m,p_T)&=&{p_T^2\over p_T^2+a^2m^2}.
\eea
While it fulfils the above condition, the matching is not exact away from the
massless limit. This is justified with the observation that the difference RS
$-$ FOM0, although formally of next-to-next-to-leading order, turns out to be
abnormally large below $p_T=5m$, so that the constant $a$ is phenomenologically
set to $a=5$. Furthermore, the FOM0 and RS results are evaluated by replacing
the transverse momentum $p_T$ with the transverse mass $m_T=\sqrt{p_T^2+m^2}$,
so that all contributions are evaluated at the same $m_T$. The central values
of the renormalisation and factorisation scale are also identified in all parts
with $m_T$.

Predictions for $c$ and $b$ quark production at the LHC with a centre-of-mass
energy of $\sqrt{s}=7$ TeV have been presented in Ref.\ \cite{Cacciari:2012ny},
where also the charm and bottom quark masses have been varied from 1.3$-$1.7
GeV and 4.5$-$5.0 GeV, respectively.
Note that FONLL predictions can only be made for inclusive heavy-quark distributions.
In particular, it is not possible to study correlations of the produced heavy quark
or hadron with other objects in the final state. The produced heavy hadron is of
course always collinear to the heavy quark, and there is also no information
on the kinematical distribution of its decay products.

\subsection{GM-VFNS}

The same restrictions also apply to the GM-VFNS calculation, which has also
been performed for inclusive distributions of heavy hadrons. It was originally
performed in the massless limit, valid at high $p_T$, and therefore includes
flavour creation, gluon splitting and flavour excitation processes. Subsequently
the calculation was improved by identifiying the previously omitted finite mass
terms through a comparison with the massive NLO calculation, where together with
the mass logarithms also finite terms were subtracted in such a way that in the
limit $m\to 0$ the correct massless $\overline{\rm MS}$ result was recovered
\cite{Kniehl:2005mk}. This is necessary, since the PDFs and FFs that are convoluted
with the partonic cross sections are defined in the ZM-VFNS. Heavy-quark mass terms
in flavour excitation processes were neglected, which corresponds to a specific
choice of scheme (known as S-ACOT), but no loss in precision \cite{Kramer:2000hn}.

Massless non-perturbative $D$-meson FFs were obtained in Ref.\ \cite{Kneesch:2007ey}
from Belle, CLEO, ALEPH and OPAL data using a starting scale of $\mu_0=m_c=1.5$ GeV
for all partons up to the charm quark and $m_b=5.0$ GeV for bottom quarks. Although
the value of $m_b$ is slightly higher here than our default value of 4.75 GeV, the
effect on charm meson production is not expected to be very large. In the
theoretical NLO calculation, finite-mass corrections were included in the production
cross sections, i.e.\ logarithmic singularities were subtracted
together with finite terms, so that the correct massless limit was recovered when
$m_q\to0$. The optimal functional ansatz at the starting scale for $c$ and $b$ quarks
fragmenting into charmed mesons turned out to be \cite{Bowler:1981sb}
\bea
 D_h(z,\mu_0)&=&N z^{-(1+\gamma^2)}(1-z)^a e^{-\gamma^2/z}
\eea
with three free parameters, which were fitted separately to $D^0$, $D^+$ and $D^{*+}$ data.
GM-FVNS predictions for $D$-meson production have previously been compared to ALICE data
obtained in pp collisions at the LHC with a centre-of-mass energy of 7 TeV
\cite{ALICE:2011aa,Kniehl:2012ti}. In the predictions for $D_s$ mesons
\cite{Kniehl:2012ti}, somewhat older FFs functions \cite{Kniehl:2006mw} had to be used.

Massless non-perturbative $B$-meson FFs were obtained in Ref.\ \cite{Kniehl:2008zza}
from ALEPH, OPAL and SLD data using a starting scale of $\mu_0=m_b=4.5$ GeV (i.e.\
slightly lower than our default value of 4.75 GeV) for the initial bottom quark FF. All
other FFs vanished there. Again, in the theoretical NLO calculation finite-mass corrections
were included in the production cross sections, but not in the FFs. A standard functional
ansatz at the starting scale
\bea
 D_h(z,\mu_0)&=&Nz^\alpha(1-z)^\beta
\eea
was found to describe the experimental data very well. In agreement with previous findings
using the FONLL approach \cite{Cacciari:2005rk} it was found that using the Peterson form
at the starting scale does not lead to a good description of the data.  Since only inclusive
$B$-meson data are available, separate fits for $B^0$ and $B^+$ mesons were not possible.
GM-VNFS predictions have been compared to inclusive $B$-meson data from CMS obtained in pp
collisions at 7 TeV centre-of-mass energy \cite{Kniehl:2011bk} and also to ALICE, ATLAS and
CMS data on leptonic decays of charm and bottom production at 2.76 and 7 TeV centre-of-mass
energy \cite{Bolzoni:2012kx}. Furthermore, predictions have been made for $D$ mesons
produced at 7 TeV in $B$ decays using either a two-step transition $b\to B\to D$, based on
$b\to B$ FFs and $B\to D$ spectra as measured by CLEO, or a one-step transition based on FFs
for $b\to D$ \cite{Bolzoni:2013vya}. Note that in the GM-VFNS approach the theoretical uncertainty
is estimated varying three independent scales ($\mu$, $\mu_f$, and $\mu_D$), not only two as in
FONLL (where $\mu_f=\mu_D$), and no fixed phenomenomenological matching function $G(m,p_T)$
is used, which generally leads to larger uncertainty bands, in particular at low $p_T$.

\subsection{POWHEG}

In contrast to the FONLL and GM-VFNS approaches, which are based on NLO calculations
and the factorisation of heavy-quark FFs and are thus limited to the description of
the inclusive production of heavy quarks and mesons, general purpose Monte
Carlo generators based on parton showers and string or cluster fragmentation
allow for a more complete description of the final state, including decay
kinematics, particle identification and, if needed, even detector response.
Their precision has now been enhanced to NLO in the hard and improved
leading-logarithm accuracy in the soft/collinear regime through a consistent
combination of NLO calculations with parton showers, made possible by the proper
subtraction of doubly counted contributions in the soft and collinear regions.

For heavy-quark production, this was first achieved in MC@NLO \cite{Frixione:2003ei},
where the subtraction terms were, however, laborious to obtain, only a matching to
the HERWIG \cite{Corcella:2000bw} parton shower was performed, and negative-weight
events could occur.
These disadvantages were subsequently improved upon in POWHEG \cite{Frixione:2007nw},
where the subtraction terms are relatively easy to calculate, a matching was
performed not only to the HERWIG, but also the PYTHIA \cite{Sjostrand:2007gs} parton shower,
and only positive event weights occur. Like MC@NLO, the POWHEG approach is based on a massive
NLO calculation, i.e.\ the massive quarks are not active partons in the PDFs and large
logarithms are not resummed into heavy-quark PDFs. Heavy flavours can, however, be
excited through the initial-state parton shower or be produced in gluon splittings.
For a correct matching of parton showers and PDFs, that is performed as in standard
(leading order) Monte Carlo programs, five-flavour strong coupling constant and PDFs
are used in MC@NLO and POWHEG, the terms $-\as \frac{2T_F}{3\pi}\log\frac{\mu^2}{m^2}\;
\sigma^{(0)}_{q\bar{q}}$ and $-\as \frac{2T_F}{3\pi}\log\frac{\mu^2}{\mu_f^2}\;\sigma^{(0)}_{gg}$
are added to the $q\bar{q}$ and $gg$ channel cross sections, and the heavy flavour
is ignored in the PDFs.
The error committed in this way is of higher order in $\alpha_s$ and numerically small.
Due to the complex colour flow in parton-parton scattering, the Sudakov form factor for
light partons in POWHEG has currently only leading-logarithmic accuracy. I.e., in contrast
to FONLL, only a subset of the large logarithms are resummed. However, at small and moderate
$p_T$ the NLO and parton shower approach should be superior to FONLL, since it has almost
the same accuracy in this region, but in addition allows for a complete and fully exclusive description
of final state. This approach should eventually also permit to include rescattering
effects in the medium in heavy-ion collisions.

In this work, the heavy-quark part of POWHEG BOX 2.1 has been called for each centre-of-mass
energy from a self-written
C++ code together with LHAPDF 5.8.9 \cite{Whalley:2005nh} and PYTHIA 8.175
\cite{Sjostrand:2007gs} to generate ten million events for both charm and bottom quark
pairs, then to shower, hadronise and decay them to stable particles. Note that the
hadronisation, performed in PYTHIA using the Lund string model, has been tuned to data at
leading, not next-to-leading order accuracy. The event files
were then analysed for heavy $D$ and $B$ mesons and their electron and muon decay
products. Kinematic (in particular rapidity) cuts were applied, and binned $p_T$
differential cross sections, normalised to the total NLO POWHEG cross section, were
calculated for every set of PDFs and choice of scale combination. The total PDF error
was obtained according to Eqs.\ (\ref{eq:a}) and (\ref{eq:b}) and stored separately
from the largest scale error.

\section{LHC predictions}
\label{sec:3}

Let us now turn to our numerical predictions for heavy-quark production in pp collisions
at the LHC with centre-of-mass energies of $\sqrt{s}=2.76$, 5.02 and 7 TeV in the ALICE kinematic
regime. In the following figures, the statistical and systematic uncertainties of
the ALICE data will always be shown as in the experimental publications as black error bars
and boxes, respectively. The theoretical
predictions will appear as shaded bands, which in the case of FONLL (green bands) comprise
variations of renormalisation scale $\mu$ and common factorisation scale $\mu_f$
as well as variations of the heavy quark mass, added in quadrature. In the case of
GM-VFNS (black dashed lines), they comprise the maximal error (yellow bands) due to
variations of the renormalisation
scale $\mu$, PDF factorisation scale $\mu_f$ and FF factorisation scale $\mu_D$, and in
the case of POWHEG (red full lines)
they comprise on the one hand the maximal error due to variations of the
renormalisation scale $\mu$ and PDF factorisation scale $\mu_f$ (light blue bands) and on the
other hand the error due to the PDF uncertainty as described above (dark blue bands).
For future comparisons of theoretical predictions with nuclear collision data it is
important to not only know the central prediction with CTEQ6.6 PDFs (red), but also
with CTEQ6.1 PDFs \cite{Stump:2003yu} (not shown), which is used as the basis of
nuclear PDFs, in particular EPS09 \cite{Eskola:2009uj}. The set CTEQ6.1 has been
obtained with the same heavy-quark masses and value of $\Lambda_{\overline{\rm MS}}^{n_f=5}$
as CTEQ6.6. We have verified that the two central predictions coincide well
in the kinematic regimes considered here.

\subsection{pp collisions at $\sqrt{s}=2.76$ TeV}

\begin{figure}
 \centering
 \epsfig{file=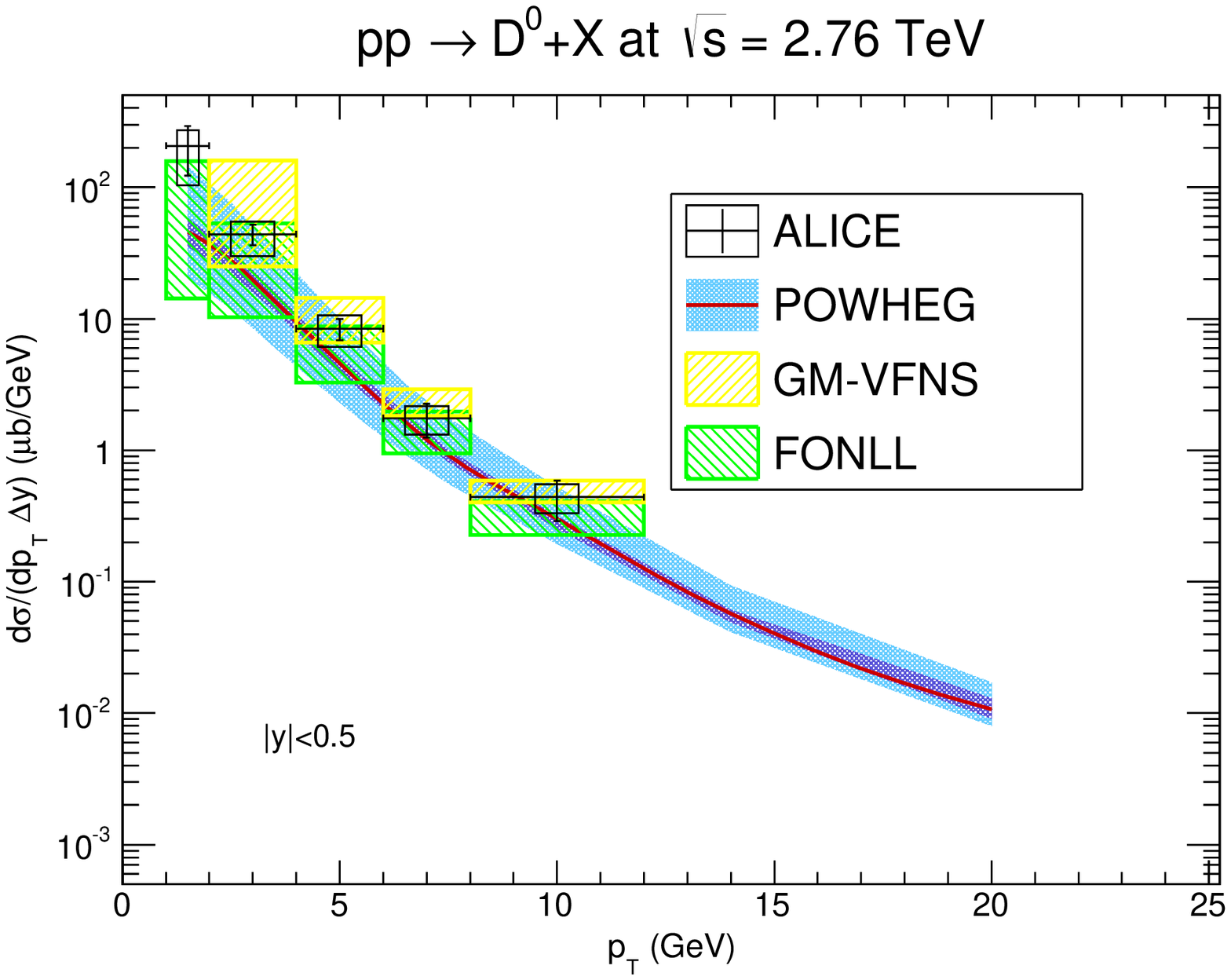,width=0.62\columnwidth}
 \epsfig{file=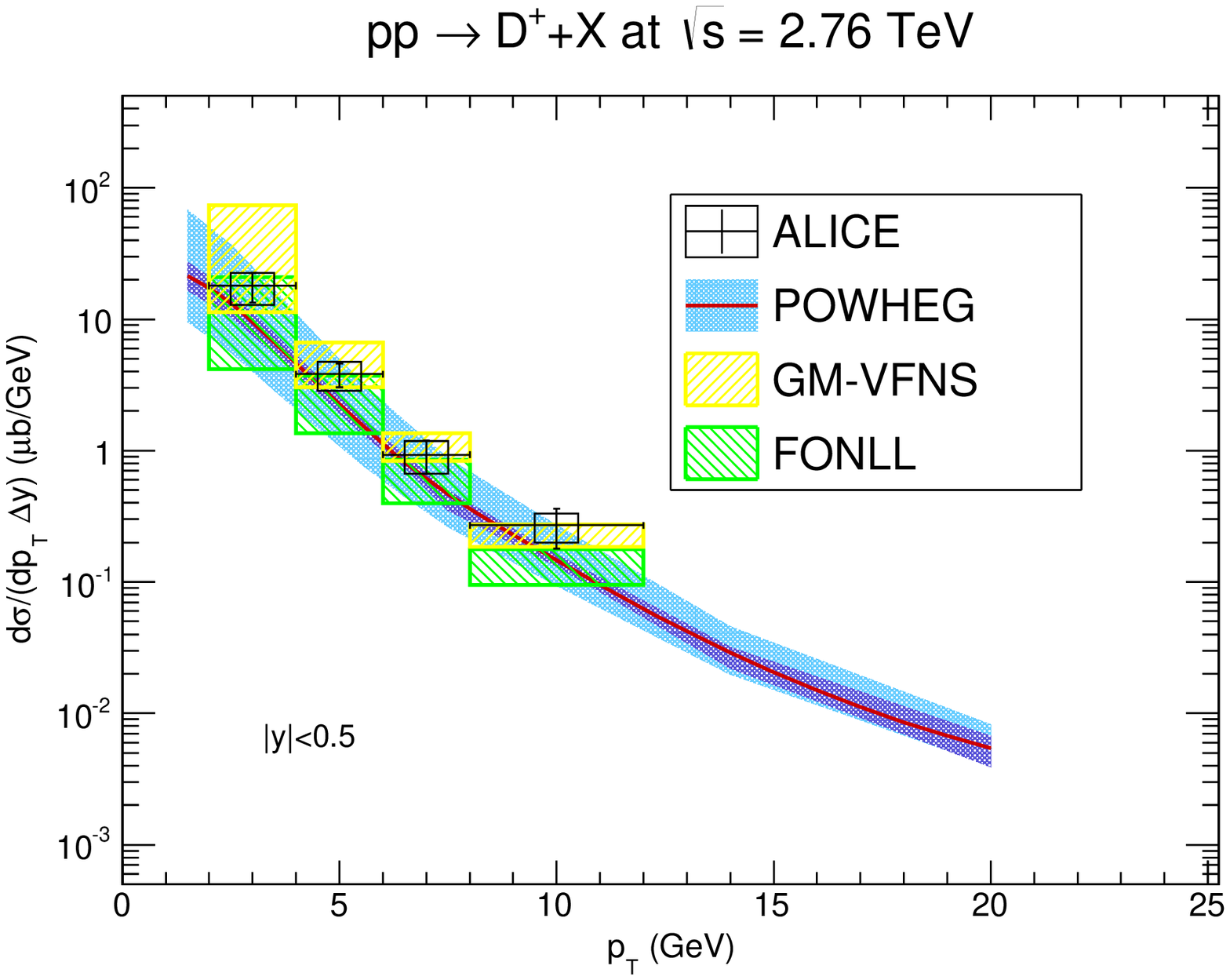,width=0.62\columnwidth}
 \epsfig{file=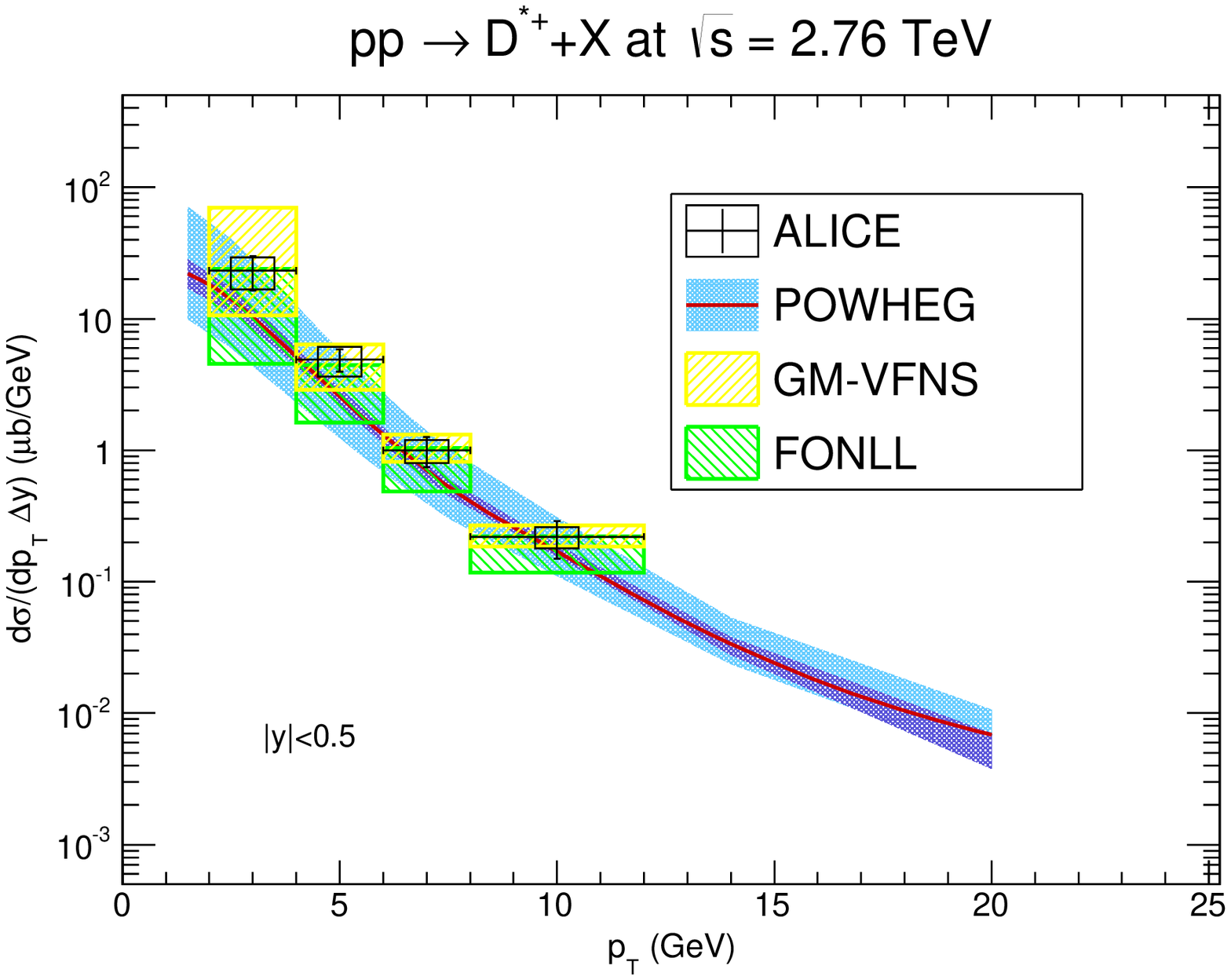,width=0.62\columnwidth}
 \caption{\label{fig:1}Transverse-momentum distributions of $D^0$ (top), $D^+$
 (centre) and $D^{*+}$ (bottom) mesons centrally produced at the LHC with
 $\sqrt{s}=2.76$ TeV and compared to ALICE data \cite{Abelev:2012vra}.}
\end{figure}

At $\sqrt{s}=2.76$ TeV, the ALICE collaboration has measured the $p_T$ distributions
of $D^0$, $D^+$ and $D^{*+}$ mesons produced centrally with rapidity $|y|<0.5$
from $p_T=1$ GeV (for $D^0$) and 2 GeV (for $D^+$ and $D^{*+}$) to 12 GeV \cite{Abelev:2012vra}.
These data are shown in Fig.\ \ref{fig:1}. Within their respective uncertainties,
%
%
which can amount to almost an order of magnitude at low $p_T$, all
three theoretical calculations agree with the data. The two originally massive
FONLL and POWHEG calculations show a tendency to underestimate the data, in
particular at low $p_T$, whereas the originally massless GM-VFNS calculation
shows the opposite tendency and tends to slightly overestimate the data, but
agrees very well with them already at intermediate to larger $p_T$ values. This
corresponds well with the expectation that in this regime, quark mass effects
should no longer play an important role. As expected from previous studies,
the scale uncertainties dominate over quark mass and PDF uncertainties, so
that the FONLL, GM-VFNS and POWHEG theoretical error bands are all comparable
in size despite their different decompositions. In particular, the PDF
uncertainty, which was computed here with POWHEG for the first time, amounts to
half or less of the scale uncertainty in the whole kinematic range, reflecting the good
knowledge of PDFs in the intermediate range of $x$ or $x_T=2p_T/\sqrt{s}=
0.001-0.02$ relevant here.

The ALICE collaboration has also analysed the production of $D^0$, $D^+$ and $D^{*+}$
mesons in PbPb collisions with $\sqrt{s_{NN}}=2.76$ TeV \cite{ALICE:2012ab}. They compared
them to scaled pp data from $\sqrt{s}=7$ TeV on the basis of the theoretical energy
dependence, as the above data set was not yet available
at the time, and established a nuclear modification factor $R_{AA}$ of about
0.3 for central and 0.6 for peripheral heavy-ion collisions. In this work, we are
only concerned with pp collisions and do not attempt to describe nuclear suppression
effects.

\begin{figure}
 \centering
 \epsfig{file=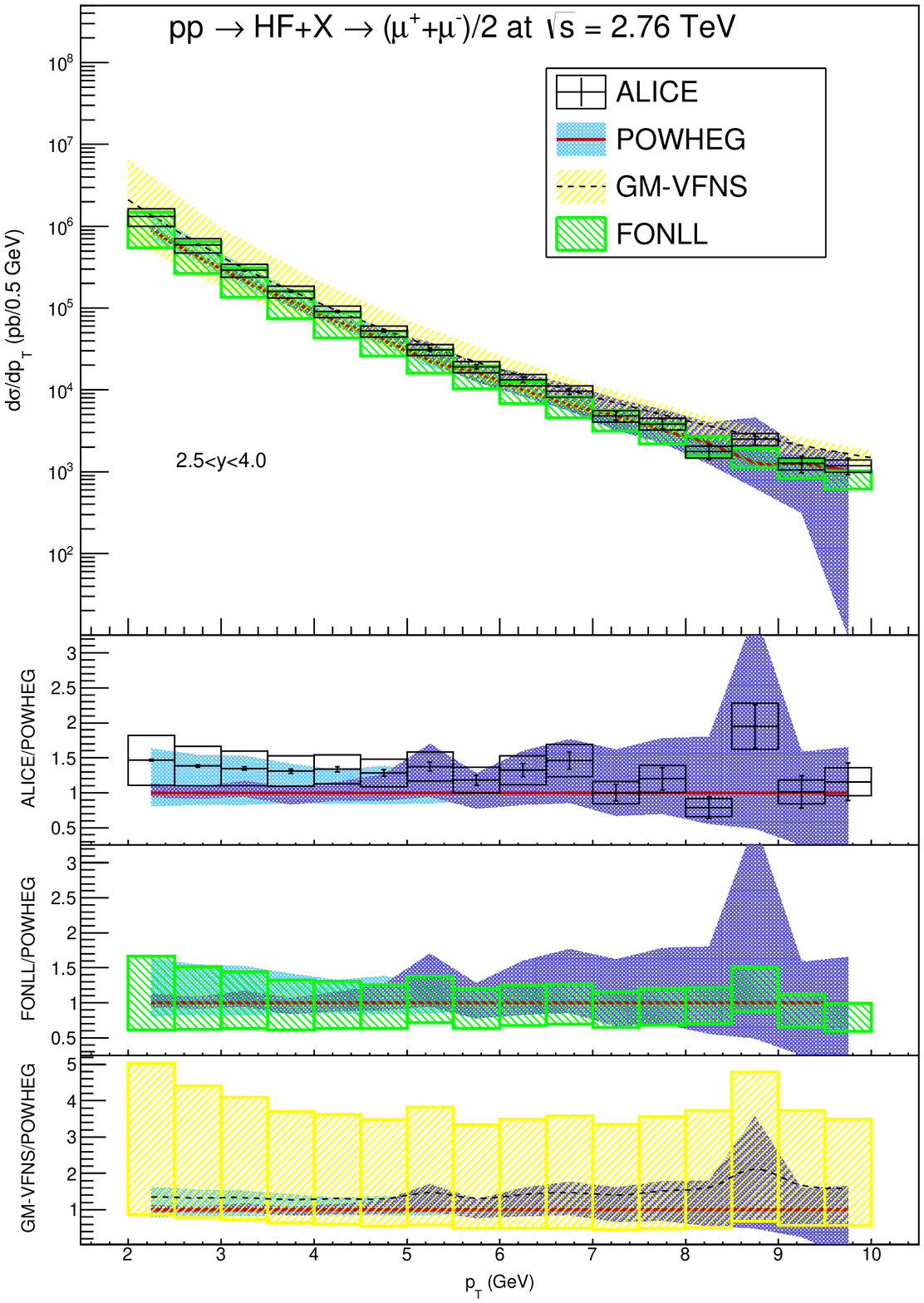,width=0.62\columnwidth}
 \caption{\label{fig:2}Transverse-momentum distributions of muons from heavy-flavour
 (charm and bottom quark) decay produced in the forward region at the LHC with $\sqrt{s}=2.76$
 TeV and compared to ALICE data \cite{Abelev:2012qh}.}
\end{figure}

The ALICE collaboration has furthermore
measured muons from heavy-flavour decay at forward rapidity
of $2.5<y<4$ in both pp and PbPb collisions \cite{Abelev:2012qh} and observed
similar nuclear suppression effects as in prompt ($B$ feed-down subtracted)
charmed meson production.
In this inclusive muon measurement, contributions from charm and bottom quarks
were not separated, but the main backgrounds from pion and kaon decays were removed.
In Fig.\ \ref{fig:2} the ALICE pp data are compared with the three theoretical
%
%
predictions in the range 2 GeV $<p_T<10$ GeV. We find again generally good
agreement and a tendency of the originally massive calculations
to slightly underestimate the data. The central GM-VFNS predictions agree
with the data perfectly well (see also Fig.\ 3 in the Erratum of Ref.\
\cite{Bolzoni:2012kx}). In the POWHEG prediction, the PDF uncertainty,
computed here for the first time, has become the dominating source of
uncertainty in this forward regime and in particular at large $p_T$.
This reflects the fact that we are now in an asymmetric situation, probing
$x$ values down to $10^{-4}$ in one proton and above 0.1 in the other. In 
both regions the PDFs are known with much less precision than at intermediate
values of $x$. Statistical Monte Carlo fluctuations can sometimes even
lead to very large deviations, in particular at large $p_T$. As one can
see from the ratio plots in Fig.\ \ref{fig:2}, the differences of ALICE
data, FONLL and GM-VFNS vs.\ POWHEG concern mostly the normalisation and
not the shape of the distributions.

\subsection{pp collisions at $\sqrt{s}=7$ TeV}

\begin{figure}
 \centering
 \epsfig{file=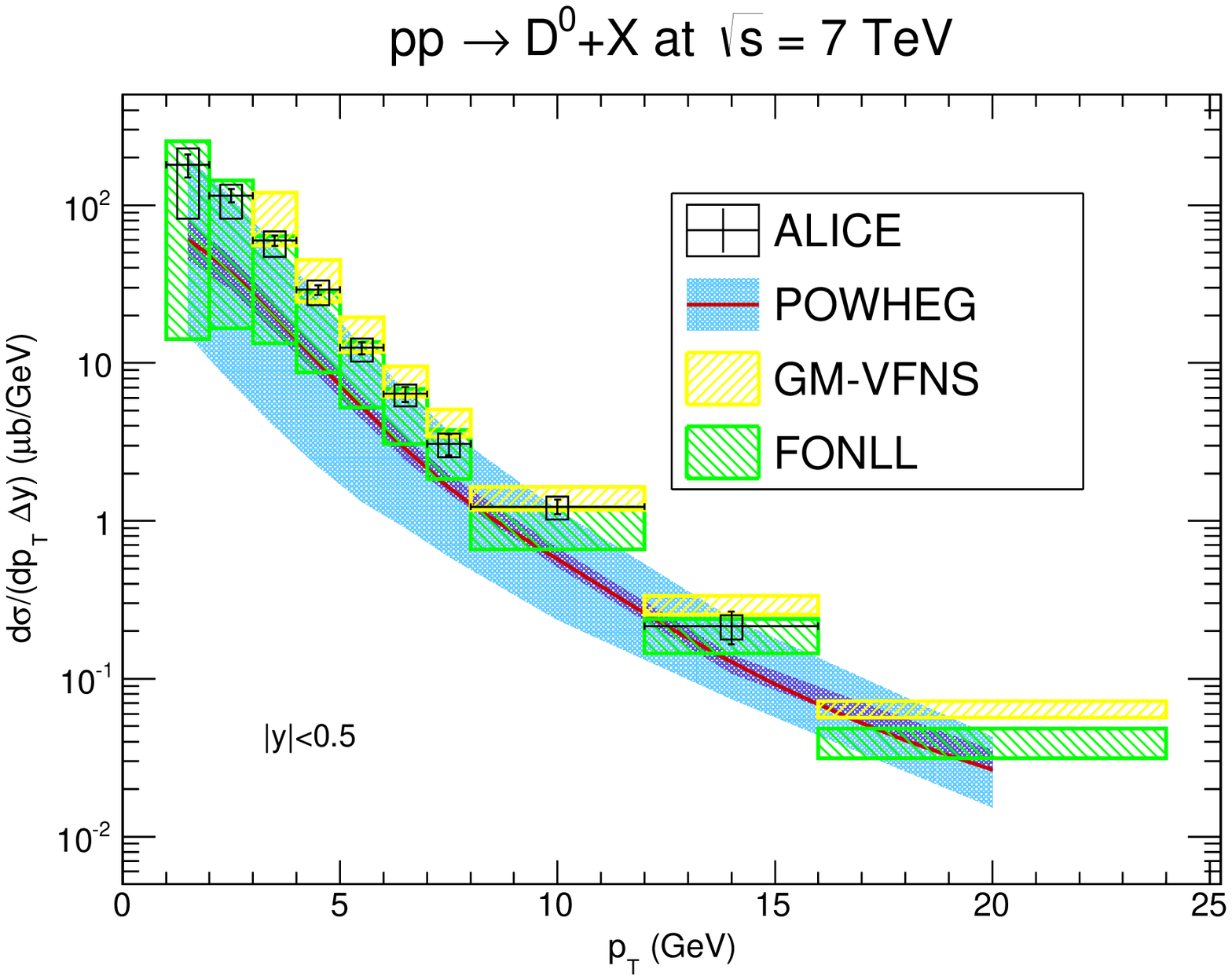,width=0.62\columnwidth}
 \epsfig{file=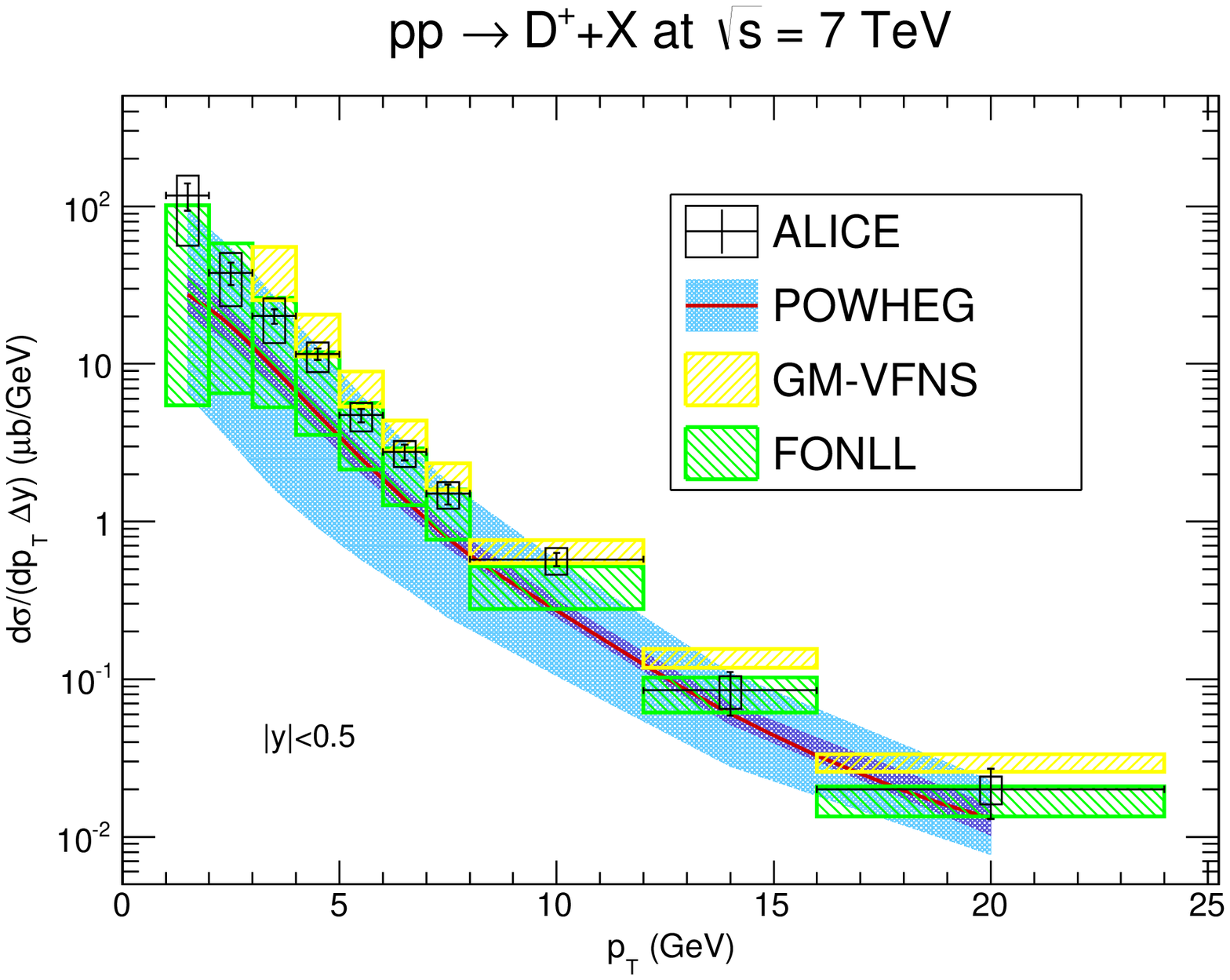,width=0.62\columnwidth}
 \epsfig{file=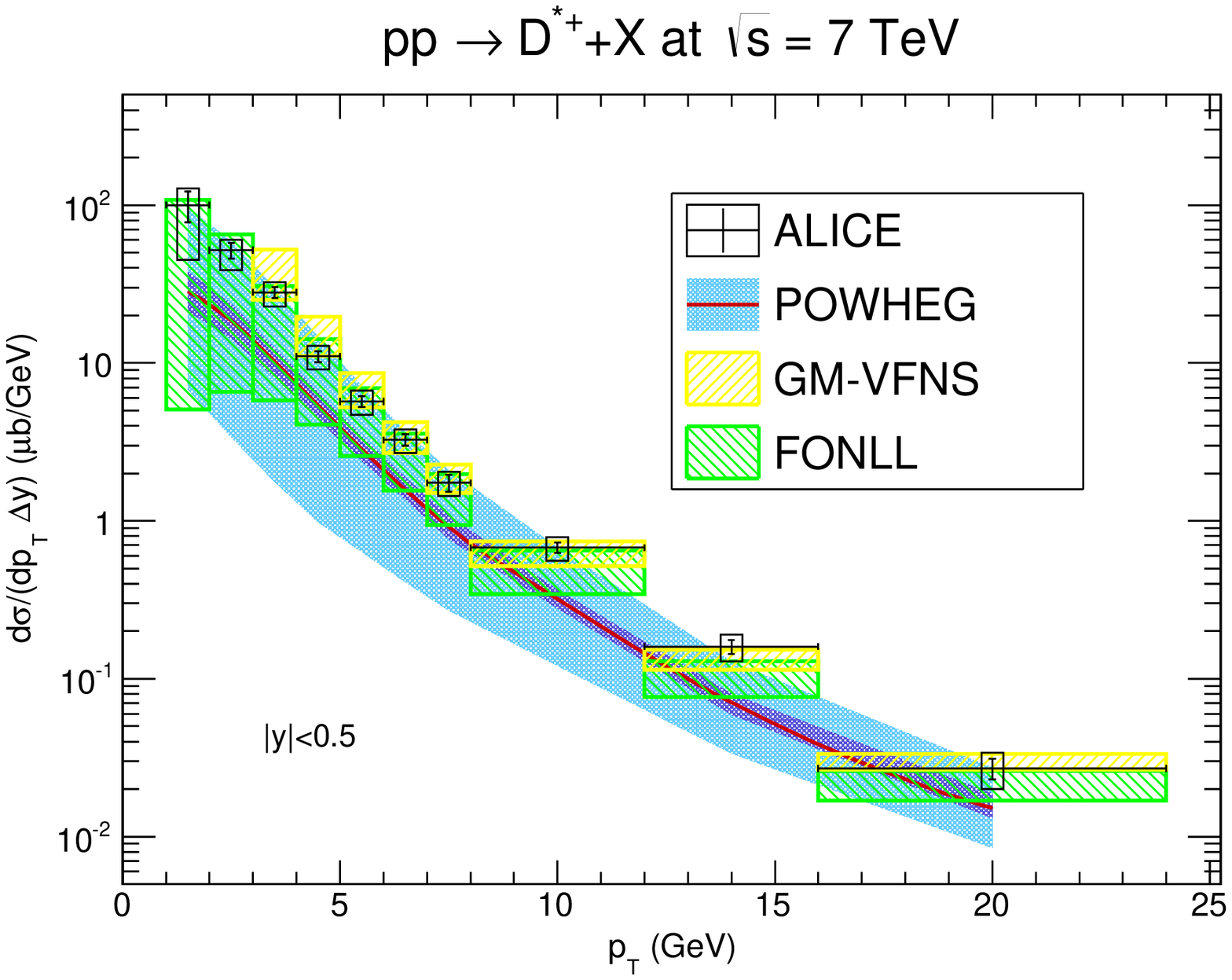,width=0.62\columnwidth}
 \caption{\label{fig:3}Transverse-momentum distributions of $D^0$ (top), $D^+$
 (centre) and $D^{*+}$ (bottom) mesons centrally produced at the LHC with
 $\sqrt{s}=7$ TeV and compared to ALICE data \cite{ALICE:2011aa}.}
\end{figure}

The prompt production of $D^0$, $D^+$ and $D^{*+}$ mesons has been analysed 
by the ALICE collaboration also at $\sqrt{s}=7$ TeV \cite{ALICE:2011aa}. As at
the lower centre-of-mass energy, the charmed mesons were identified through their
decays $D^0\to K^-\pi^+$, $D^+\to K^-\pi^+\pi^+$, $D^{*+}\to D^0\pi^+$ and their
charge conjugates. Feed-down from $B$-meson decays was subtracted using FONLL.
The corresponding $p_T$-distributions at central rapidity $|y|<0.5$ are shown in
Fig.\ \ref{fig:3}. The discussion from the previous section applies here again
%
%
in the sense that within the considerable theoretical uncertainties
the data can be described in all three approaches. The main difference
is that the data extend now out to $p_T$ values of 16 GeV for $D^0$
and 24 GeV for $D^+$ and $D^{*+}$. At large $p_T$,
where mass effects become less important, the GM-VFNS prediction
shows the largest stability, followed by FONLL, which resums
mass effects into a perturbative FF at next-to-leading logarithmic level,
whereas in POWHEG
these logarithms are only partially resummed through the Sudakov factor
in the parton shower. At low $p_T$, only the FONLL and POWHEG predictions
are shown to agree with the data, since the GM-VFNS central prediction
and its scale uncertainty rise there rapidly and in particular do not
show the specific turnover of the data as the genuinely massive calculations
\cite{Kniehl:2012ti}. \footnote{We take the opportunity to correct the misprints
in the figure captions of Ref.\ \cite{Kniehl:2012ti}, where the centre-of-mass
energy should read $\sqrt{s}=7$ TeV, not GeV. It has been shown that for a
specific combination of scales, the GM-VFNS predictions can be brought into
agreement with the data also at low $p_T$ \cite{Spiesberger:2012zn}.} As before,
the PDF uncertainties remain subdominant over the whole kinematic range.

\begin{figure}
 \centering
 \epsfig{file=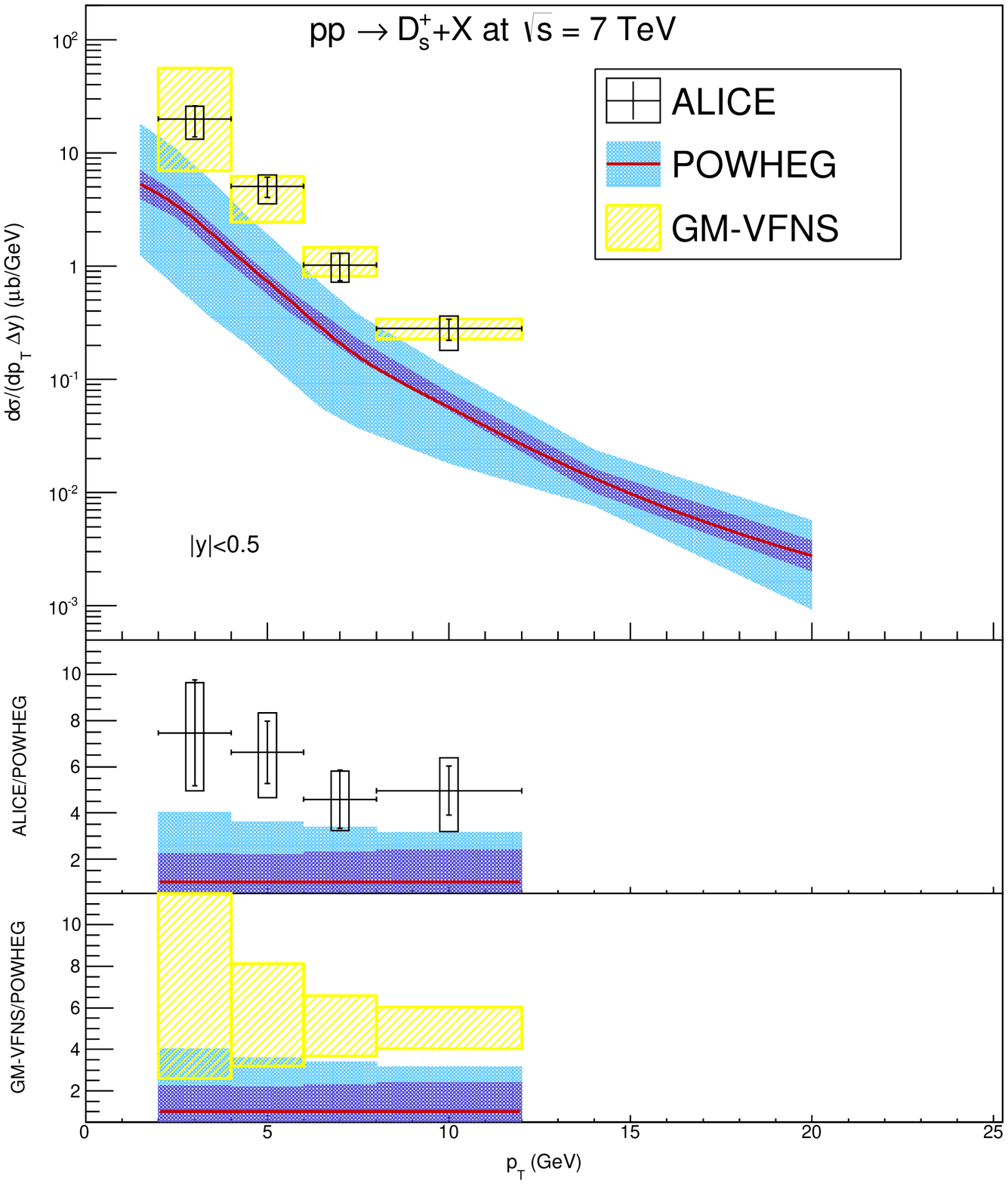,width=0.62\columnwidth}
 \caption{\label{fig:4}Transverse-momentum distributions of $D_s^+$ mesons centrally
 produced at the LHC with $\sqrt{s}=7$ TeV and compared to ALICE data \cite{Abelev:2012tca}.}
\end{figure}

For centrally produced $D_s^+$ mesons, which were reconstructed by ALICE through the
decay $D_s^+\to \phi\pi^+$, with $\phi\to K^-K^+$, and its charged conjugate
\cite{Abelev:2012tca}, the $p_T$-spectrum extends only from 2 to 12 GeV, as can
be seen in Fig.\ \ref{fig:4}. The GM-VFNS predictions agree here very well with
%
%
the data, whereas POWHEG falls short of them even when taking into account its
uncertainty band. This may be seen as an indication that the mass of the strange
quark leads to a further suppression of the predicted rate and/or
that the fragmentation into bound states of charm and strange quarks is less well
described by the Lund string model than the one into heavy-light mesons and may
thus require more tuning to data. As the ratio plots in Fig.\ \ref{fig:4} show,
the ALICE data and GM-VFNS predictions differ from POWHEG mostly in normalisation,
but also somewhat in shape.

\begin{figure}
 \centering
 \epsfig{file=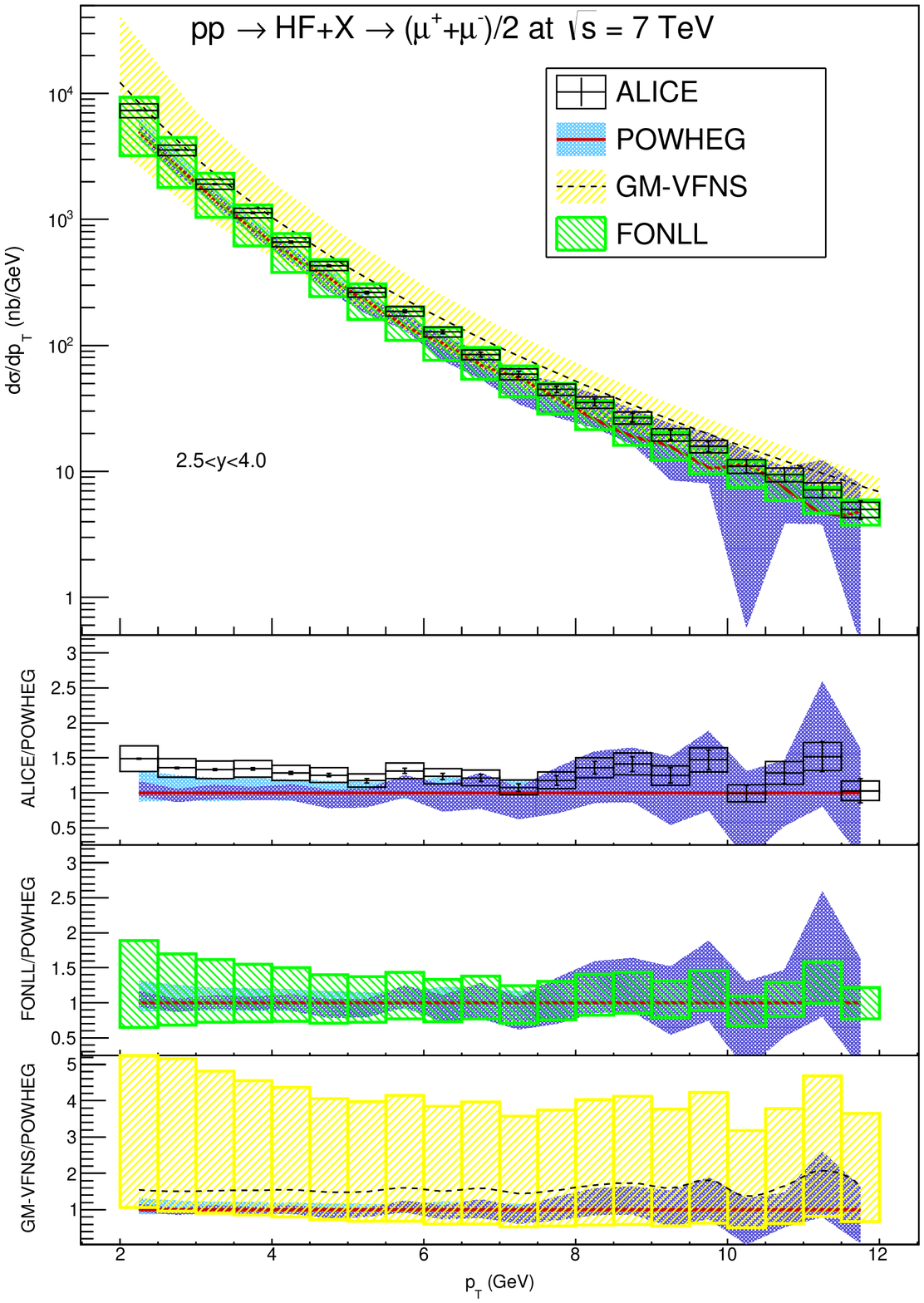,width=0.62\columnwidth}
 \caption{\label{fig:5}Transverse-momentum distributions of muons from heavy-flavour
 (charm and bottom quark) decay produced in the forward region at the LHC with $\sqrt{s}=7$
 TeV and compared to ALICE data \cite{Abelev:2012pi}.}
\end{figure}

Similarly to the measurements at $\sqrt{s}=2.76$ TeV, ALICE has measured muons
produced from heavy-flavour decay without flavour separation and after subtraction
of the backgrounds from pion and kaon decays in the forward region $2.5<y<4$
\cite{Abelev:2012pi}.
These data are shown in Fig.\ \ref{fig:5} together with 
%
%
the three different theoretical predictions in the range 2 GeV $<p_T<$ 12 GeV.
The central GM-VFNS prediction lies somewhat higher than at the lower centre-of-mass
energy in particular at high $p_T$ (see also Fig.\ 3 in the Erratum of Ref.\
\cite{Bolzoni:2012kx}), but within the considerable scale uncertainty the GM-VFNS
predictions still agree with the ALICE data, as do the two other
predictions from FONLL and POWHEG. The PDF uncertainties, computed only with POWHEG,
exceed those from the scale variations at intermediate and high $p_T$,
where again very small and large values of $x$ are probed, respectively.
The ALICE data, FONLL and
GM-VFNS predictions have in addition been divided into five equidistant rapidity bins and
successfully compared there, but we refrain here from showing the corresponding
figures and POWHEG predictions as they do not add significant information.

\begin{figure}
 \centering
 \epsfig{file=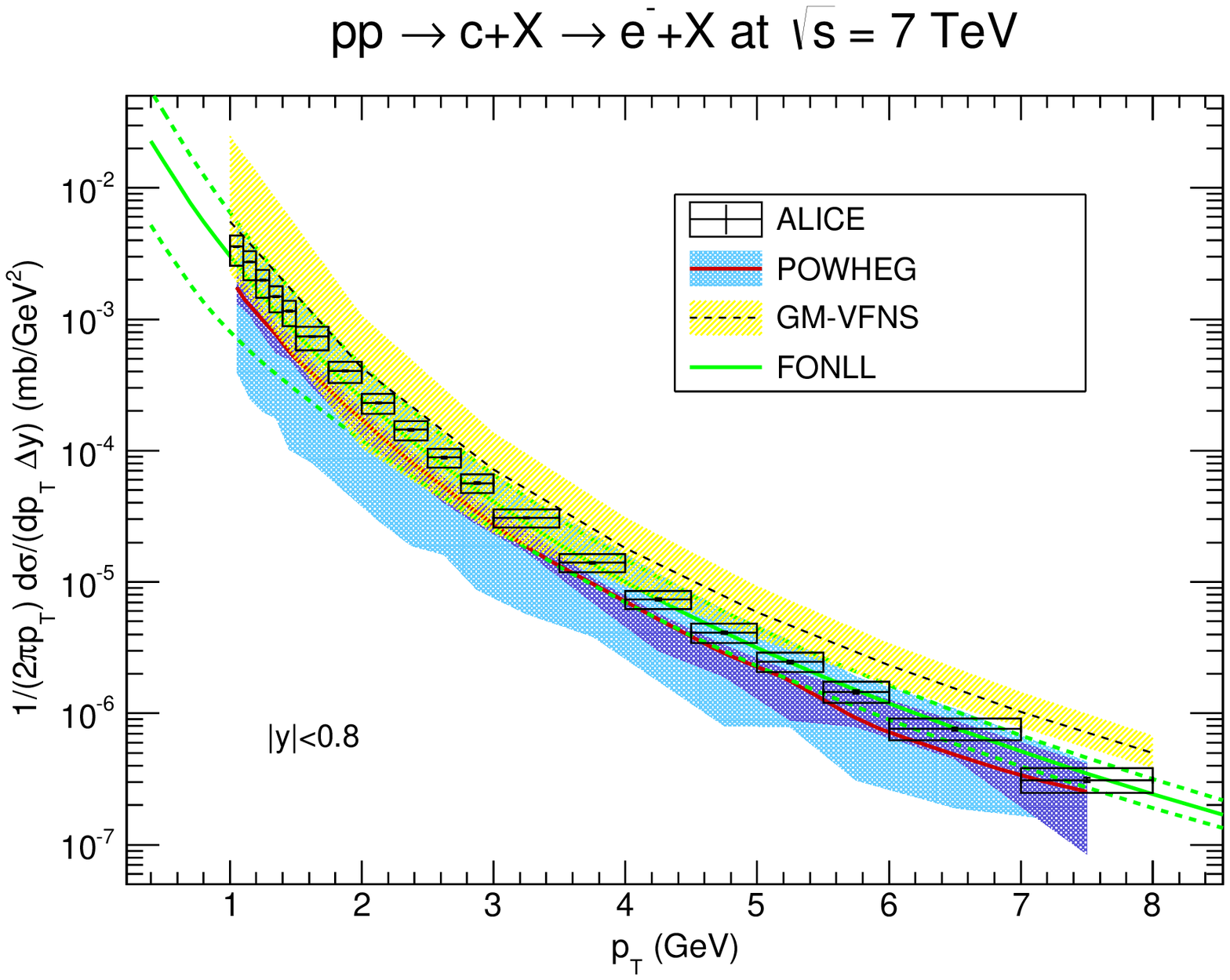,width=0.62\columnwidth}
 \epsfig{file=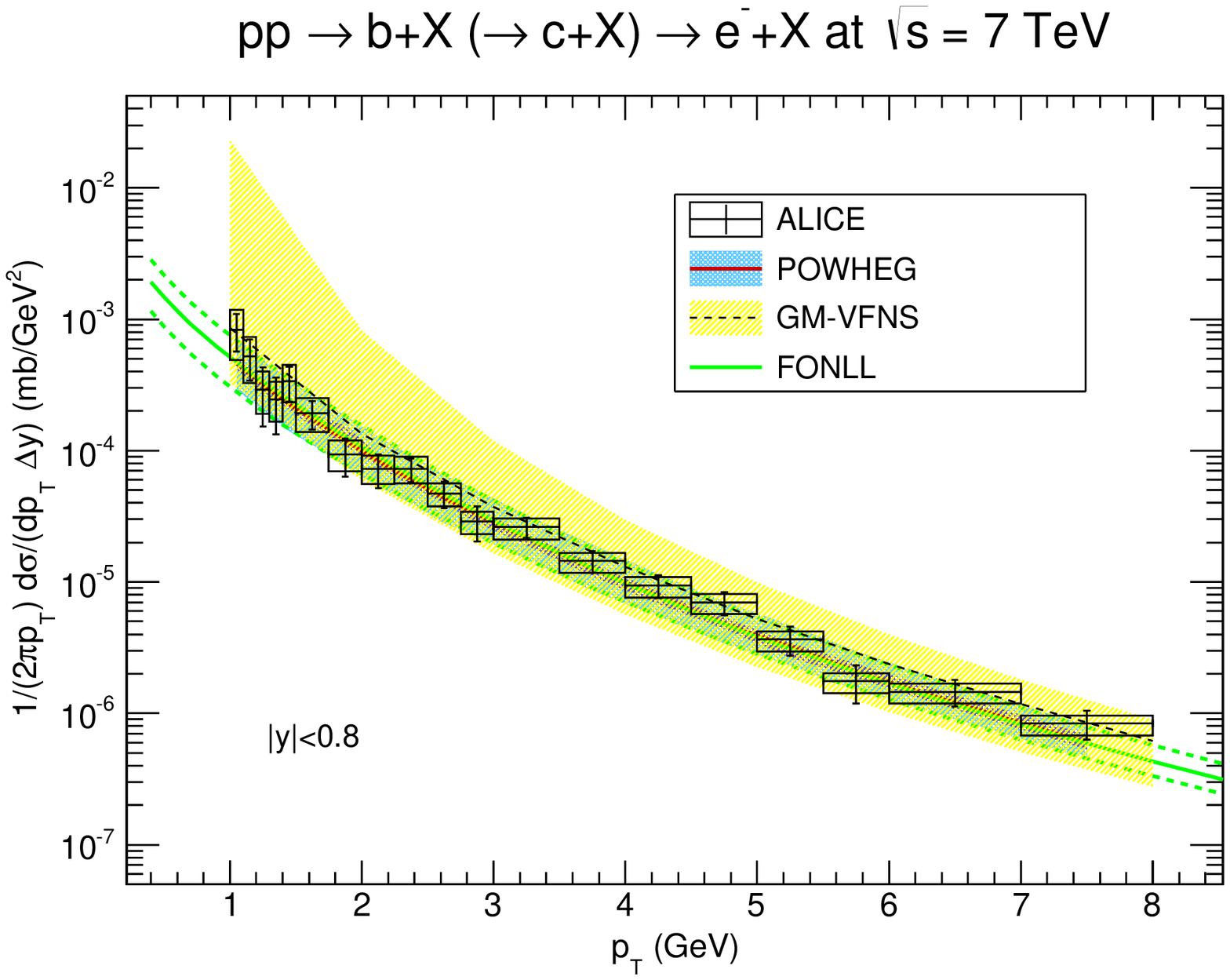,width=0.62\columnwidth}
 \caption{\label{fig:6}Transverse-momentum distributions of electrons from charm (top)
 and bottom (bottom) quark decay centrally produced at the LHC with $\sqrt{s}=7$
 TeV and compared to ALICE data \cite{Abelev:2012sca}.}
\end{figure}

At central rapidities ($|y|<0.5$), ALICE has furthermore measured heavy-flavour
decay into electrons without flavour separation \cite{Abelev:2012xe}. The main
backgrounds here stem from pseudoscalar, light and heavy vector meson decays, which
have been subtracted, together with real and virtual photon conversions, using
a Monte Carlo ``cocktail'' calculation \cite{Abelev:2012xe}.
A comparison with FONLL predictions is
included in the experimental publication, while a comparison with GM-VFNS
predictions can be found in Fig.\ 3 in the Erratum of Ref.\ \cite{Bolzoni:2012kx}.
The measurement was subsequently repeated including flavour separation, where
decays of beauty hadrons were identified through a secondary vertex, displaced
from the primary collision vertex \cite{Abelev:2012sca}.
For this data set, comparisons with FONLL have been made in the experimental
publication and with GM-VFNS in Ref.\ \cite{Bolzoni:2012kx}, but only for the
decays of bottom hadrons. As one can see in Fig.\ \ref{fig:6} (bottom), the
%
%
theoretical uncertainty for the latter is very large at small $p_T$, whereas
it is much smaller for charm decays, as can also be seen in Fig.\ \ref{fig:6} (top)
and as it should
be for smaller quark masses. For beauty decays, the POWHEG prediction and its
theoretical uncertainty coincide almost exactly with the FONLL predictions over the
entire $p_T$ range, the PDF uncertainty being again subdominant in this central
kinematic regime. For charm decays, only the central POWHEG prediction and its
upper uncertainty band limit coincide with FONLL, the lower edge being somewhat
lower. In this case, the PDF uncertainty becomes visible and comparable to,
albeit still smaller than the scale error at larger $p_T$. The excellent
agreement among
FONLL and POWHEG is indeed quite remarkable and much better for inclusive leptons
than for inclusive mesons, which obviously depend much more on the fragmentation model
than the decay leptons.
 
\subsection{pp collisions at $\sqrt{s}=5.02$ TeV}

\begin{figure}
 \centering
 \epsfig{file=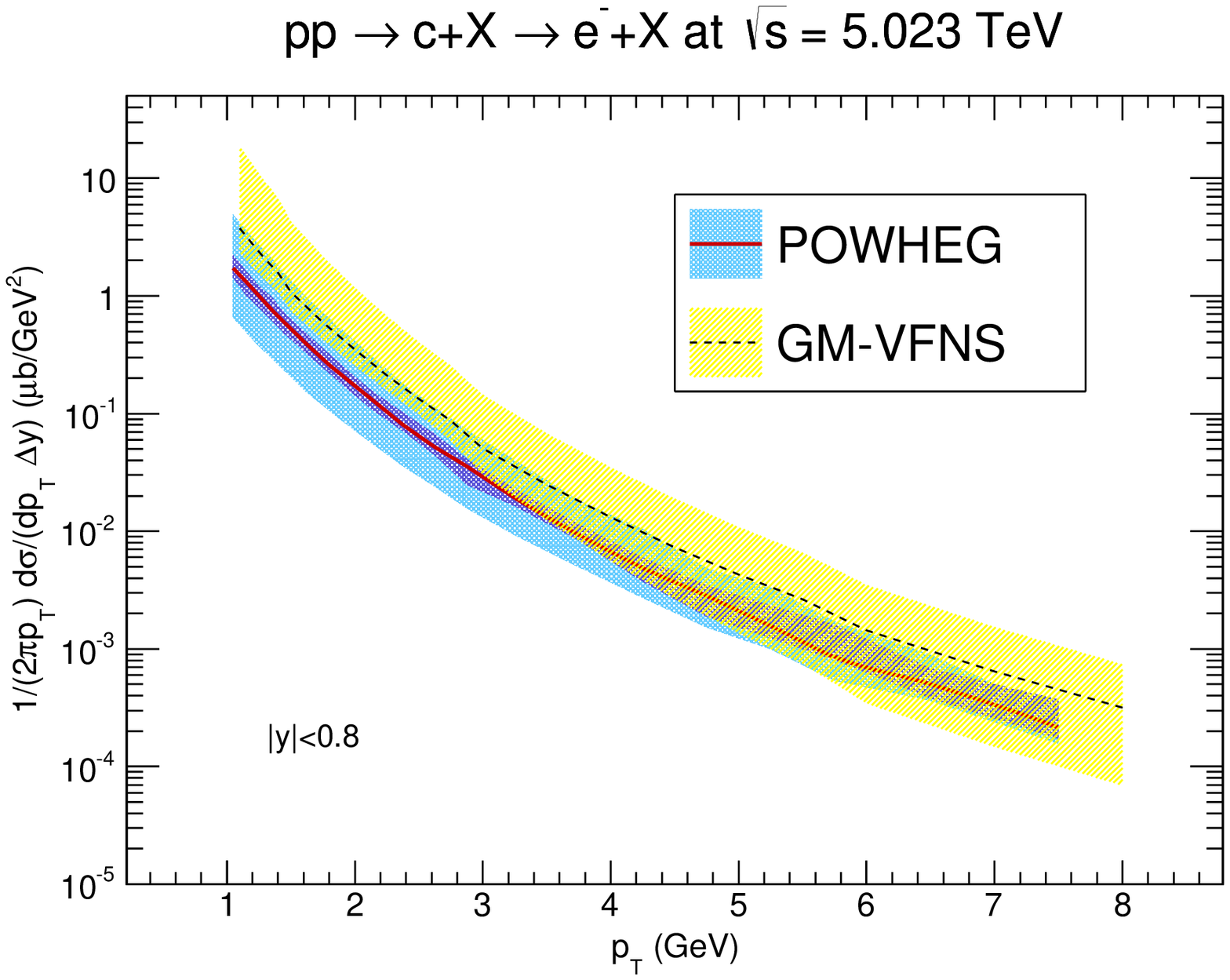,width=0.62\columnwidth}
 \epsfig{file=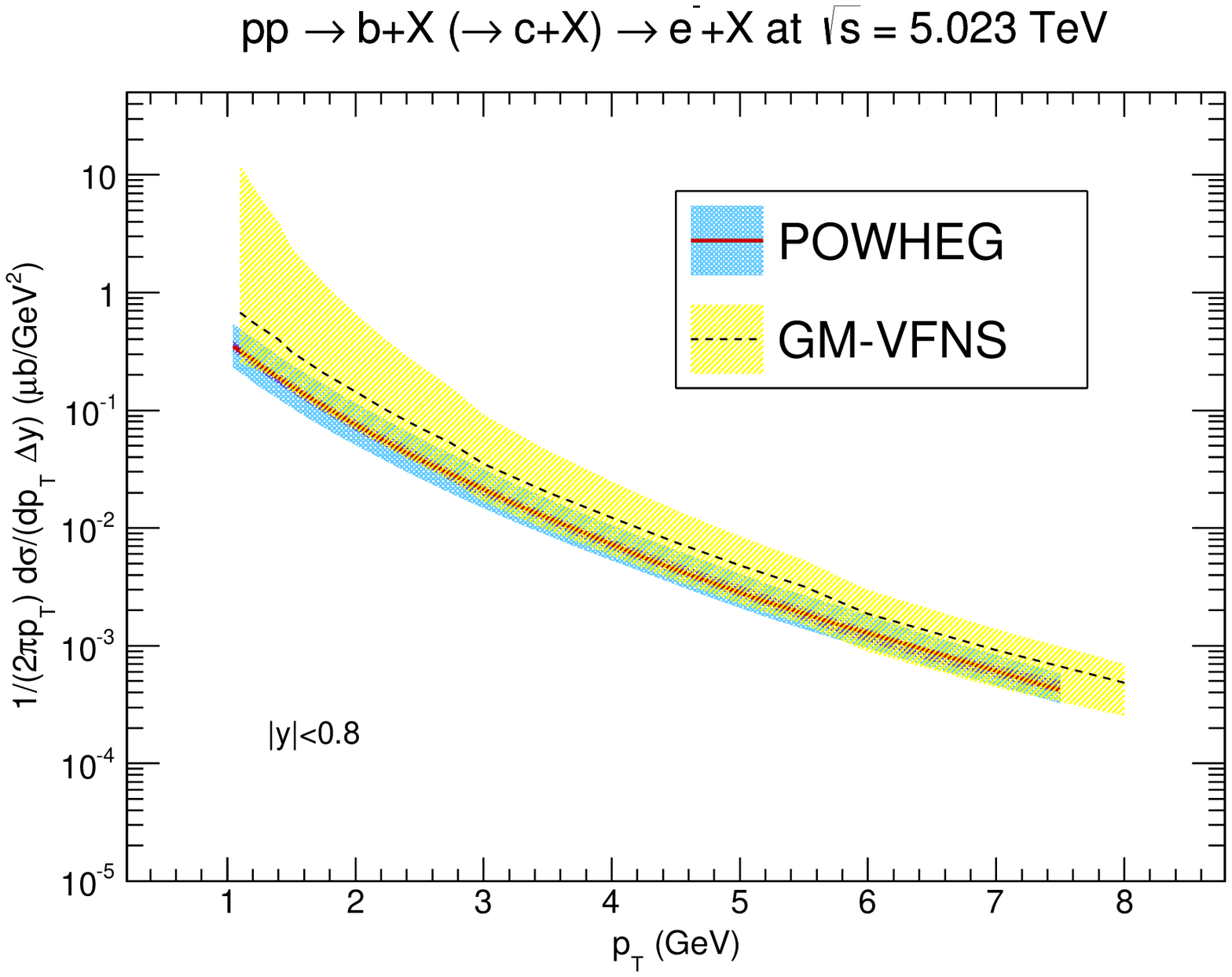,width=0.62\columnwidth}
 \caption{\label{fig:7}Transverse-momentum distributions of electrons from charm (top)
 and bottom (bottom) quark decay centrally produced at the LHC with $\sqrt{s}=5.02$
 TeV, pertinent to pPb collisions during 2013.}
\end{figure}

Finally, we turn to pp collisions at a centre-of-mass energy of $\sqrt{s}=5.02$ TeV,
relevant also for pPb collisions, where no reference calculations are published yet. In
Fig.\ \ref{fig:7} we show new predictions for centrally produced electrons from
%
%
heavy-flavour decays with POWHEG and GM-VFNS in the kinematic
regime that is currently under analysis by the ALICE collaboration, i.e.\ for
transverse momenta from 1 GeV to 8 GeV. As before, the general trend of good
agreement within scale uncertainty bands, subdominant PDF uncertainties and a
tendency of GM-VFNS to lie above POWHEG, in particular at low $p_T$, continues
here. As before, the GM-VFNS scale uncertainty at low $p_T$ is larger for
beauty than for charm hadrons due to the larger bottom quark mass.
It will be interesting to learn if the ALICE data in pPb also agree with both
theoretical predictions as was the case for pp reactions at higher and lower
energies.

\section{Conclusions}
\label{sec:4}

In this paper, motivated by the importance of a solid theoretical
understanding of the pp baseline for future analyses of heavy-ion
collision data, we have presented an extensive theoretical analysis of
heavy-quark production in the ALICE kinematic regime at the LHC. In addition
to the originally massive FONLL and massless GM-VFNS calculations, which
were partially already available, but which only allow for limited comparisons
of selected inclusive meson or decay lepton distributions, we have performed
detailed calculations with the NLO Monte Carlo program POWHEG. These allow
in principle for a more exclusive description of the final state,
including e.g.\ correlations of the heavy quarks with other particles,
and eventually also for the modelling of medium modifications.

In line with our goal of establishing a pp baseline, we have concentrated
here on detailed comparisons of the FONLL, GM-VFNS and POWHEG approaches
to inclusive transverse-momentum spectra of heavy mesons or decay leptons,
at central and forward rapidities, and at three different centre-of-mass
energies. Within the respective theoretical uncertainties, which were
defined slightly differently in the three cases, but which were always
dominated by scale variations, we found good agreement among the three
theoretical calculations. For centrally produced electrons, the agreement
among FONLL and POWHEG turned out to be indeed quite remarkable, while
the hadronisation model affected more the inclusive meson spectra,
in particular for $D_s^+$ mesons.

PDF uncertainties were analysed for the first time and obtained here
with the POWHEG approach with the result that they become dominant
only in the forward regime and/or at large transverse momenta, corresponding
to asymmetric situations and the regimes of very small and very large $x$.
There, the PDFs are still known with insufficient precision and could
be determined better using heavy-quark pp data, but e.g.\ also data
from vector boson production \cite{Brandt:2013hoa,Brandt:2014vva}.
A better knowledge of proton and nuclear PDFs will in particular be important
to distinguish cold from hot nuclear effects and to understand
central vs.\ peripheral collisions and individual vs.\ collective phenomena.

Clearly this work is only a first step towards more detailed theoretical
investigations, which will include further differential distributions,
two-particle correlations, proton-nucleon and nucleon-nucleon collisions,
and ultimately collective phenomena. Our extensive NLO Monte Carlo data
sample produced with POWHEG provides a solid basis for such studies,
where often only the analysis routines must be adapted, whereas the generated
data can be used for multiple purposes.

\acknowledgments

We thank the M.\ Heide and H.\ Spiesberger for making several previously
obtained FONLL and GM-VFNS results available to us.
The work of C.\ Klein-B\"osing was supported by the Helmholtz Alliance
Program of the Helmholtz Association, contract HA216/EMMI ``Extremes of
Density and Temperature: Cosmic Matter in the Laboratory''.




\end{document}